# The specific form of histograms presenting the distribution of data of α-decay measurements appears simultaneously in the moment of New Moon in different points from Arctic to Antarctic.

S. E. Shnoll[1,2], K. I. Zenchenko[2], S. N. Shapovalov[3], E. S. Gorshkov[3], A. V. Makarevich[3] and O. A. Troshichev[3]

[1]Lomonosov State University, Physical Department, Moscow, 119899, Russia
[2]Institute of Theoretical and Experimental Biophysics, Russian Academy of Sciences, Pushchino, Moscow Region, 142290, Russia
[3]Arctic and Antarctic Institute, 38 Beringa street, S-Peterburg, 199397, Russia
*shnoll@iteb.ru*

**Abstract.** It is shown that fine structure of distributions of fluctuations in different processes depends on position of the Earth with respect of the Moon. The same form of histogram was observed at New Moon moment independently of geographic position of laboratory and local time. Possible origin of the phenomenon is discussed.

**1 Introduction**
Even precise measurements produce scattered results. Traditionally, this scattering is viewed as an undesirable phenomenon. When scattered results are analyzed, peaks and troughs in their distribution (histogram) is commonly neglected, because they are assumed to result from random fluctuations. Commonly accepted procedures for the statistical treatment of the results of measurements involve smoothing of this fine structure and approximating histograms by Gaussian or Poisson distributions [1-3]
However, we have shown that the fine structure of histograms representing distributions of the results of measurements of a variety of processes, from biochemical reactions to radioactive decay, is not random and depends on cosmophysical factors [4- 13]. The shape of these histograms is a manifestation of the fundamental properties of our world.
An adequate object for investigating these properties is radioactive decay, a process that *a fortiori* is not subject to trivial "terrestrial" influences. In view of this, we studied, in the course of the last 20 years, the regularities of changes in the shape of histograms representing the results of measurements of *α*-decay of $^{239}$**Pu** specimens rigidly mounted on semiconductor detectors. The results of long-term "round-the-clock" measurements of the number of acts of radioactive decay per unit time have been deposited in computer archives. We constructed histograms from nonoverlapping consecutive fragments of the time series. To facilitate visualization, the histograms were smoothed by moving summation. When shapes of histograms are compared, their extension and compression along the abscissa axis, as well as and rotation about the vertical axis (mirror image), are allowed. These procedures are performed using a software tool [8].
A study of the time series from which the histograms were built by conventional mathematical methods indicates that, as it might be expected, radioactive decay is a quite random process, which obeys the Poisson statistics. However, the shape of the histograms is not random. Comparison of the shapes of histograms in many thousands of pairwise combinations revealed the following patterns:
   (1) **"Near-zone effect"** - histograms of a particular shape are often observed successively;

(2) **"Cosmophysical periodicity"** - histograms of a particular shape often appear with periodicity of 24 h ("solar day"), 23 h 56 min ("sidereal day"), approximately 27 days, and 365 days;

(3) **"Synchronous on local time scale"** - histograms of similar shapes constructed from the results of independent measurements of diverse processes (from biochemical reactions to radioactive decay) performed at remote geographical locations, are often observed at the same local time.

However extravagant these results may look, they do not contradict the basic concepts of physics. The general picture of the world has been formed as a result of studies of changes in measured quantities. Our studies deal with changes in the fine structure of distributions of the results of these measurements. These changes are of universal character and are not related to the scale of energy transformation in a particular process.

Here we show that the fine structure of histograms constructed from the measurements of the alpha-activity of $^{239}$Pu and the gamma-activity of $^{137}$Cs is largely determined by the mutual positions of the Earth, the Moon, and the Sun. The effect is most pronounced during the New-Moon period.

**2 Results**

Fig. 1 - Fig. 8 presents a series of 21 histograms, each constructed from the results of 60 2-s measurements of the alpha-activity of $^{239}$Pu specimens rigidly fixed on a semiconductor detector performed in different laboratories in different geographical points during eight New-Moon periods in 2000 and 2001. Each histogram corresponds to a 2-min interval. On the abscissa of each histogram is the radioactivity in imp/2s. On the ordinate is the frequency of occurrence of this activity value. The mean radioactivity is 870 imp/2s. Histograms of the shape characteristic of the New Moon period (called hereinafter "New-Moon histogram") are presented in the middle of each series (shown in red). For convenience of visual comparison, the histograms were smoothed 15 times by moving summation [8-10,12]. The numbers on the figures denote the ordinal numbers of successive histograms.

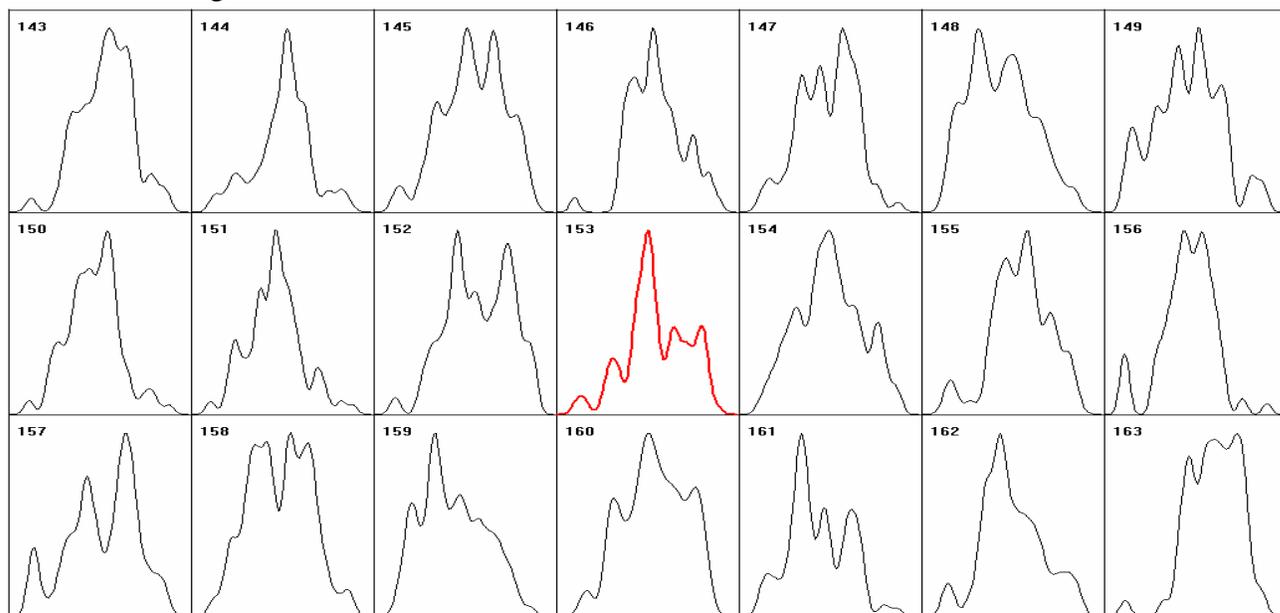

**Fig. 1** Series of histograms constructed from the measurements of alpha-radioactivity of a $^{239}$Pu specimen rigidly fixed on a semiconductor detector. Measurements were done on 31 July, 2000 in Pushchino (in latitude $54^0 50$' North, in longitude $37^0 38$' East). The New Moon was at 3h 39min (Greenwich). Histogram no. 150 corresponds to this time. The "New-Moon histogram" in this case has number 153 (is shown in red), i.e., was observed 6 min later.

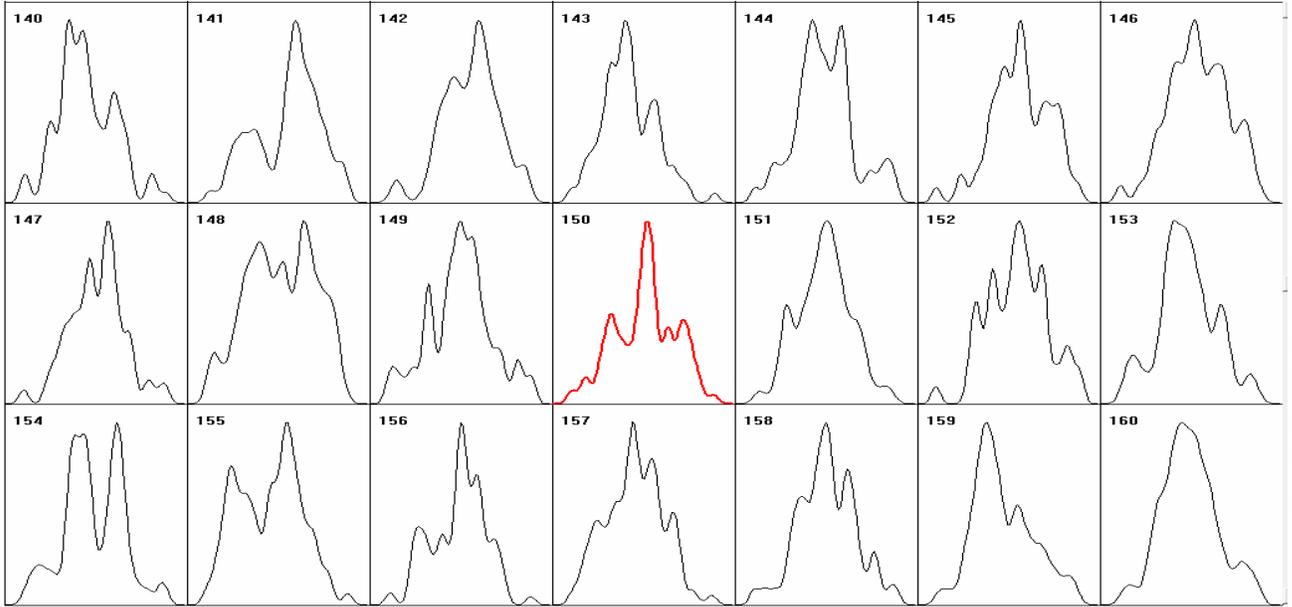

**Fig. 2** Series of histograms constructed from the measurements of alpha-radioactivity of a $^{239}$**Pu** were done on 29 August, 2000 in Pushchino. The New Moon was at 11h 21min. Histogram no. 150 corresponds to this time. The "New-Moon histogram" in this case has number 150, i.e., was observed exactly at the New Moon.

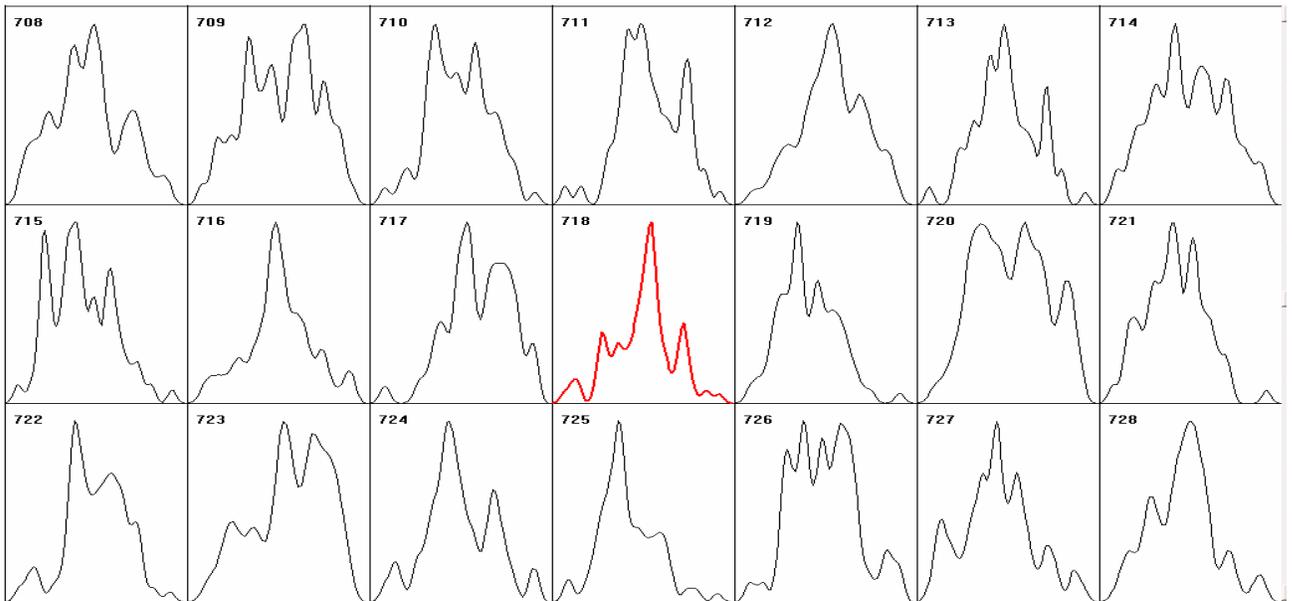

**Fig. 3** Measurements were done on 27 September 2000 on the ship "Academik Fedorov" during the Arctic expedition in the Arctic Ocean (in latitude $82^0$ North, in longitude $50^0$ East). The New Moon was at 20h 54min. Histogram no. 717 corresponds to this time. The "New-Moon histogram" in this case has number 718, i. e., was observed 2 min later.

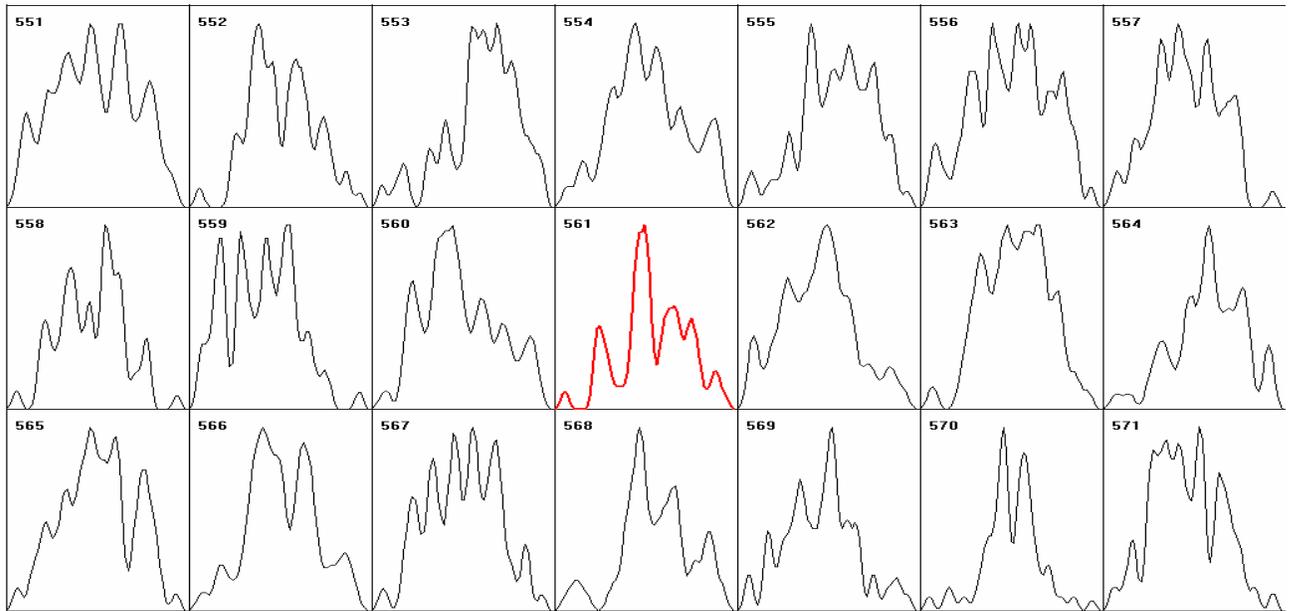

**Fig. 4** Measurements were done on 23 May 2001 on the ship "Academik Fedorov" during the Antarctic expedition (in latitude $62^0$ South, in longitude $88^0$ East). The New Moon was at 2h 48min. Histogram no. 555.5 corresponds to this time. The "New-Moon histogram" in this case has number 561, i. e., is observed 11 min later.

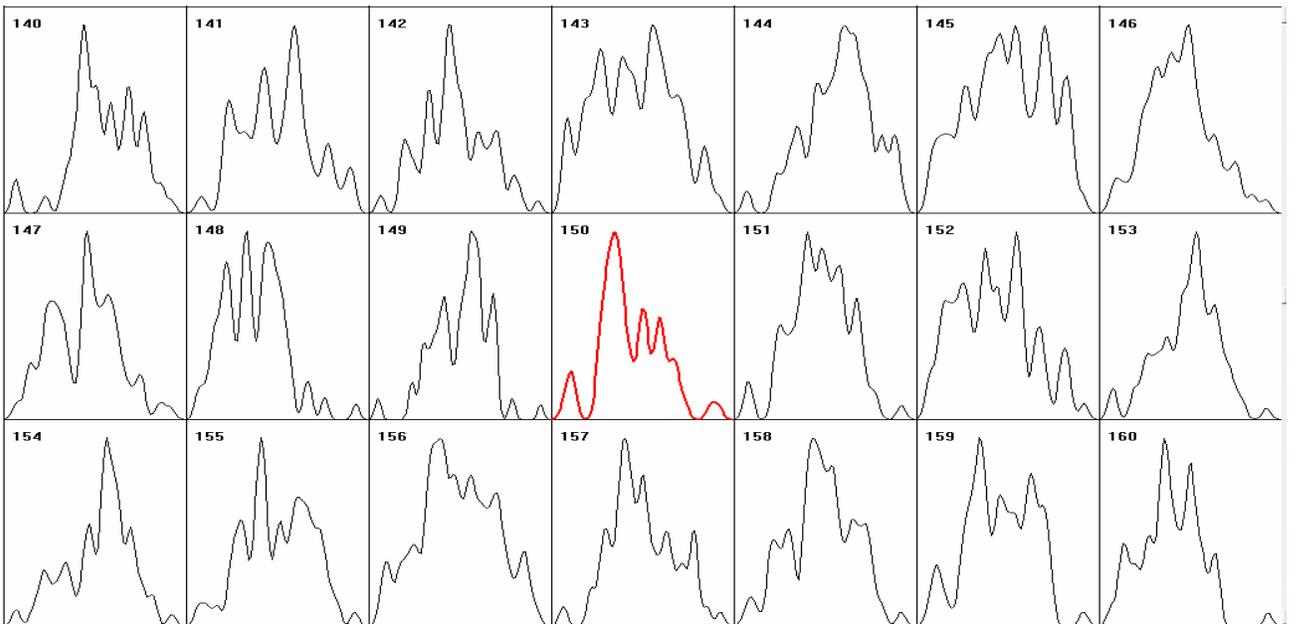

**Fig. 5** Measurements were done on 23 February, 2001 in Pushchino. The New Moon was at 8h 23min. Histogram no. 150 corresponds to this time. The "New-Moon histogram" in this case has number 150, i. e., is observed exactly at the New Moon.

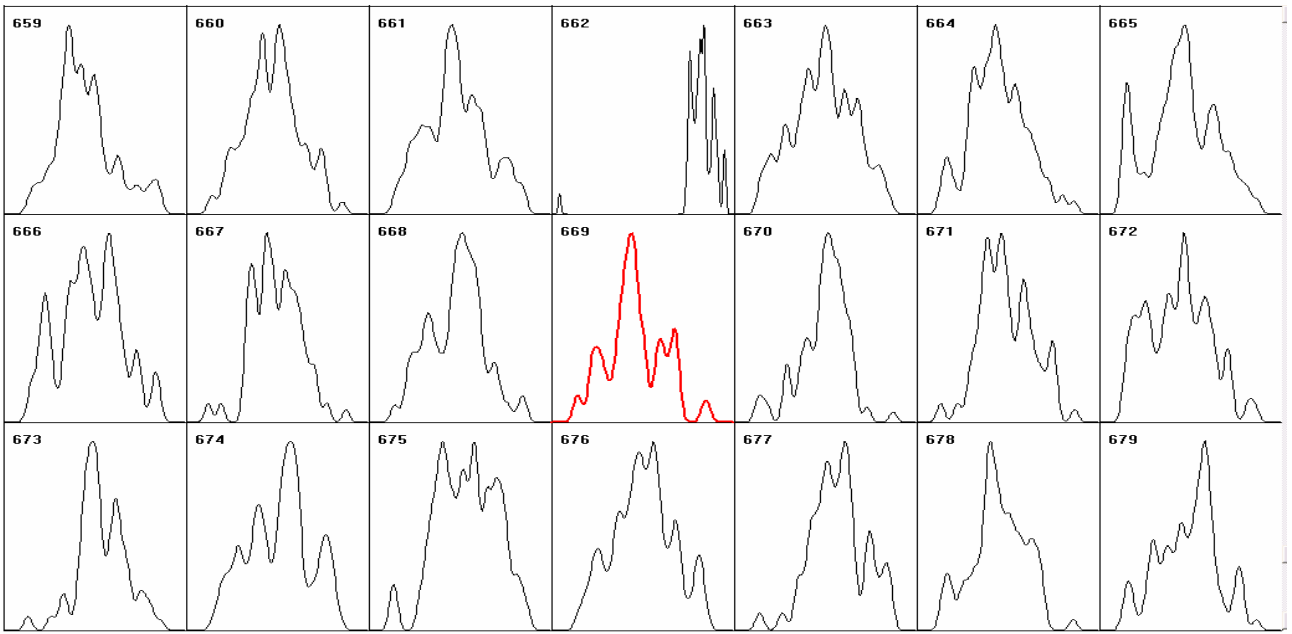

**Fig. 6** Measurements were done on 21 June, 2001 in Pushchino. The New Moon was at 11h 59min. Histogram no. 670,5 corresponds to this time. The "New-Moon histogram" in this case has number 669, i. e., is observed 3 min earlier.

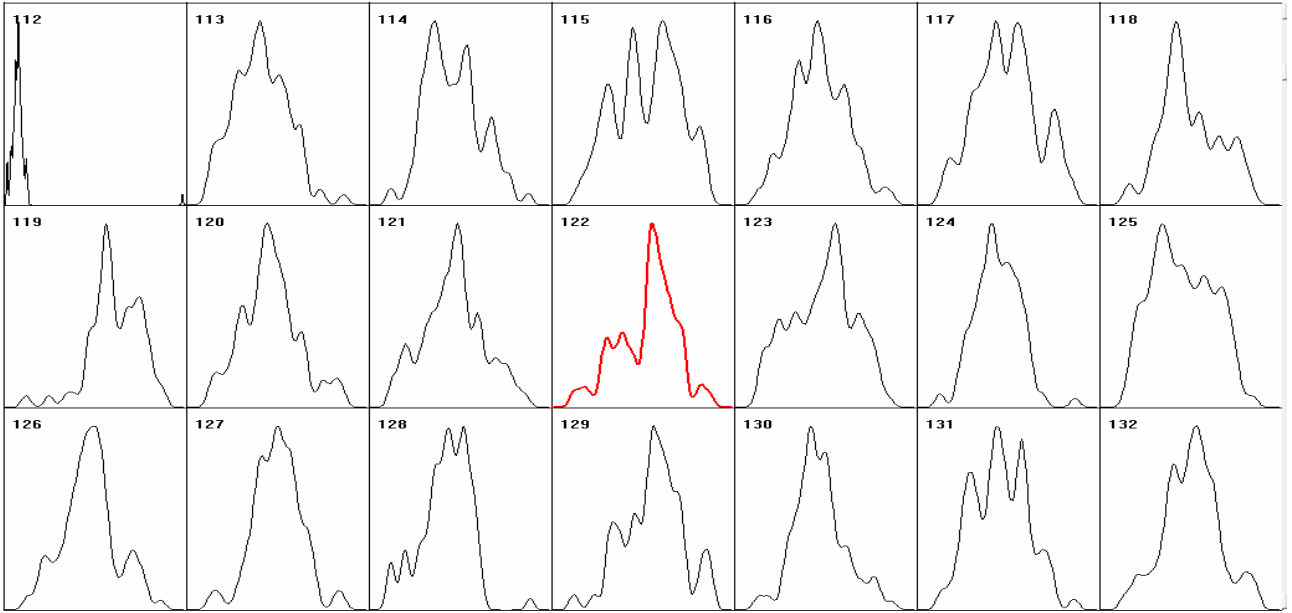

**Fig. 7** Measurements were done on 21 June 2001 on the ship "Academik Fedorov" during the Antarctic expedition (in latitude $33^0$North, in longitude $13^0$ West). The New Moon was at 2h 59min. Histogram no. 118.5 corresponds to this time. The "New-Moon histogram" in this case has number 122, i. e., is observed 5 min later.

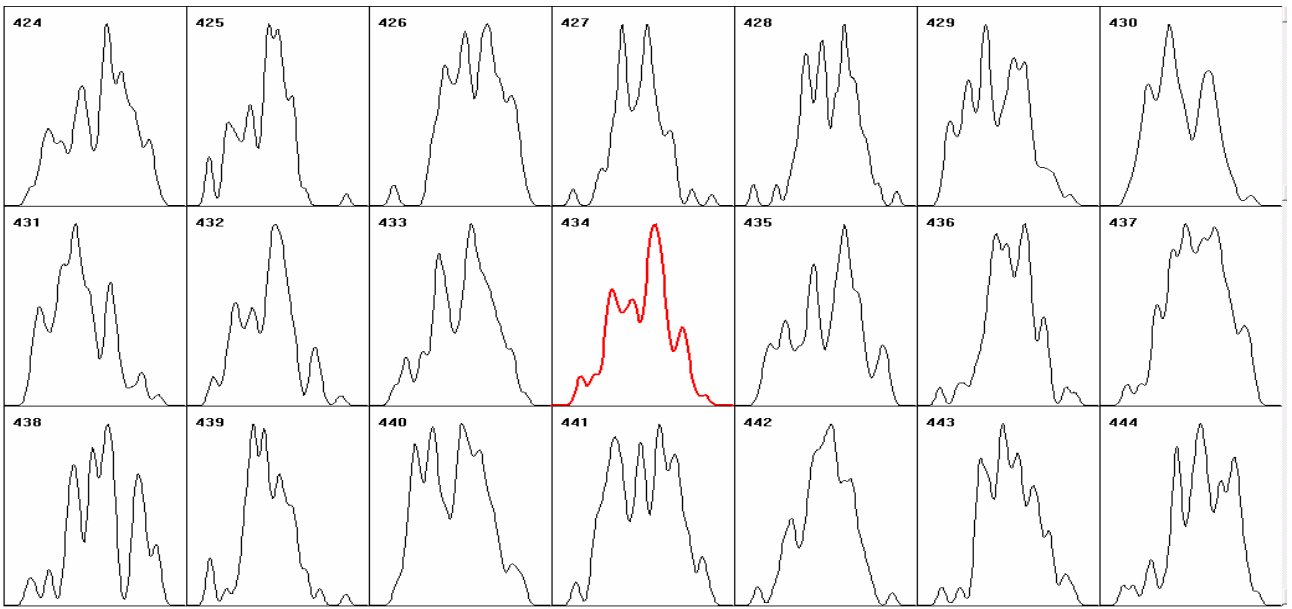

**Fig. 8** Measurements were done on 17 September 2001 in Pushchino. The New Moon was at 10h 28min. Histogram no. 434 corresponds to this time. The "New-Moon histograms" in this case have numbers 432 and 434, i. e., are observed 4 min earlier and exactly at the New Moon.

It is evident from this series of figures that, during the New Moon, "New-Moon histogram" is observed almost simultaneously at different geographical points, regardless of the season and the position of the Moon relative to the horizon at a given geographical point.

Superposition of histograms on a series of Fig. 9 - Fig. 13 clearly demonstrates a similarity of the histograms at the onset of the New Moon period at different geographical points.

Fig. 9 shows a superposition of the histograms constructed from the results of measurements of $^{239}$Pu alpha-radioactivity in Pushchino at the onset of the New Moon period on July 31, 2000 (no. 153) and on June 21, 2001 (no. 669). Their similarity illustrates a uniformity of histogram forms at the New Moon at different times; the interval in this case was 11 months.

Fig. 10 shows a synchronism in the occurrence of single-type histograms upon measurements of alpha-radioactivity of $^{239}$Pu in Pushchino (in latitude $54^0 50'$ North, in longitude $37^0 38'$ East) (no. 669) and on the ship "Academik Fedorov" during the Antarctic expedition (in latitude $33^0$ South, in longitude $13^0$ West) (no. 122) at New Moon on 21 June 2001.

Fig. 11 shows a synchronism in the occurrence of similar histograms upon measurements of alpha-radioactivity of $^{239}$Pu in Pushchino (in latitude $54^0 50'$ North, in longitude $37^0 38'$ East) (no. 147) and on the ship "Academik Fedorov" during the Arctic expedition (in latitude $80^0$ North, in longitude $50^0$ East) (no. 718) at the New Moon on September 27, 2000.

Fig. 12 shows a similarity in the form of histograms upon measurements of alpha-radioactivity of $^{239}$Pu in Pushchino (in latitude $54^0 50'$ North, in longitude $37^0 38'$ East) (no. 434) at the New Moon on September 17, 2001 and on the ship "Academik Fedorov" during the Antarctic expedition (in latitude $63^0$ South, in longitude $88^0$ East) (no. 561) at the New Moon on May 23, 2001.

Fig. 13 shows a similarity in the form of histograms upon measurements of alpha radioactivity of $^{239}$Pu at the New Moon on September 27, 2000 on the ship "Academik Fedorov" during the Arctic expedition (in latitude $80^0$ North, in longitude $50^0$ East) (no. 718) and on the ship "Academik Fedorov" during the Antarctic expedition (in latitude $33^0$ South, in longitude $13^0$ East) (no. 122) at the New Moon on June 21, 2001.

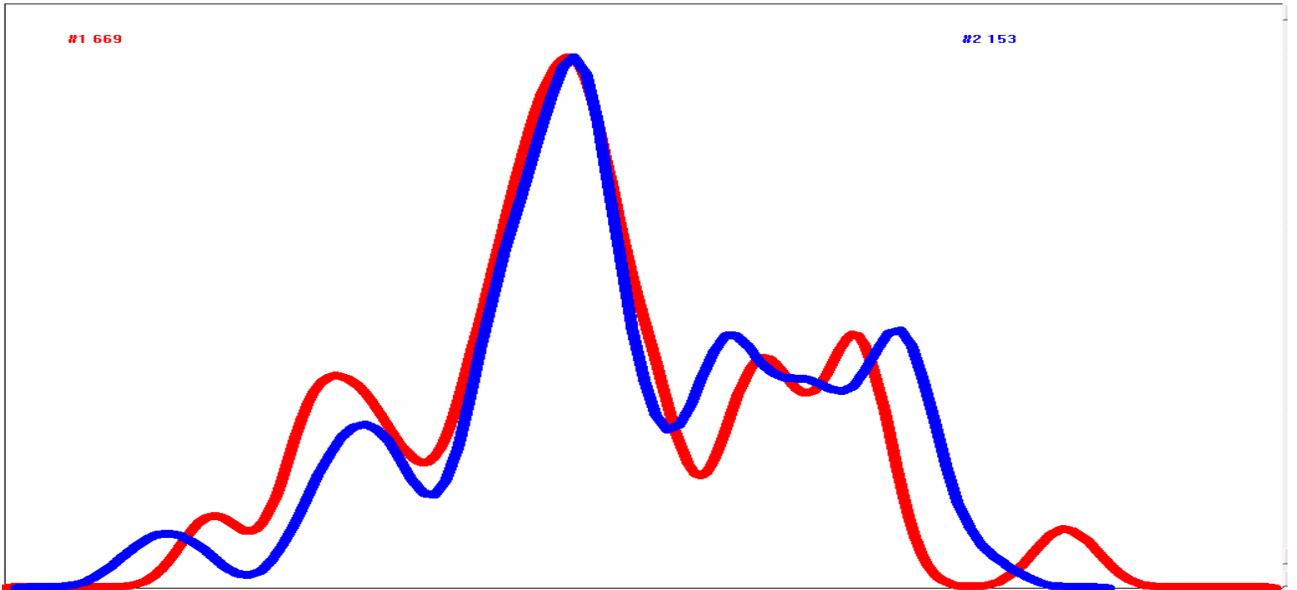

**Fig. 9** A similarity of the New-Moon histograms **at the same geographic point at different times** (Pushchino in latitude $54^0 50'$ North, in longitude $37^0 38'$ East, the New Moon, 31 July 2000, histogram no. 153 and the New Moon on 21 June 2001, no. 669). Each histogram was constructed from the results of sixty 2-s measurements of of $^{239}$**Pu** alpha-decay.

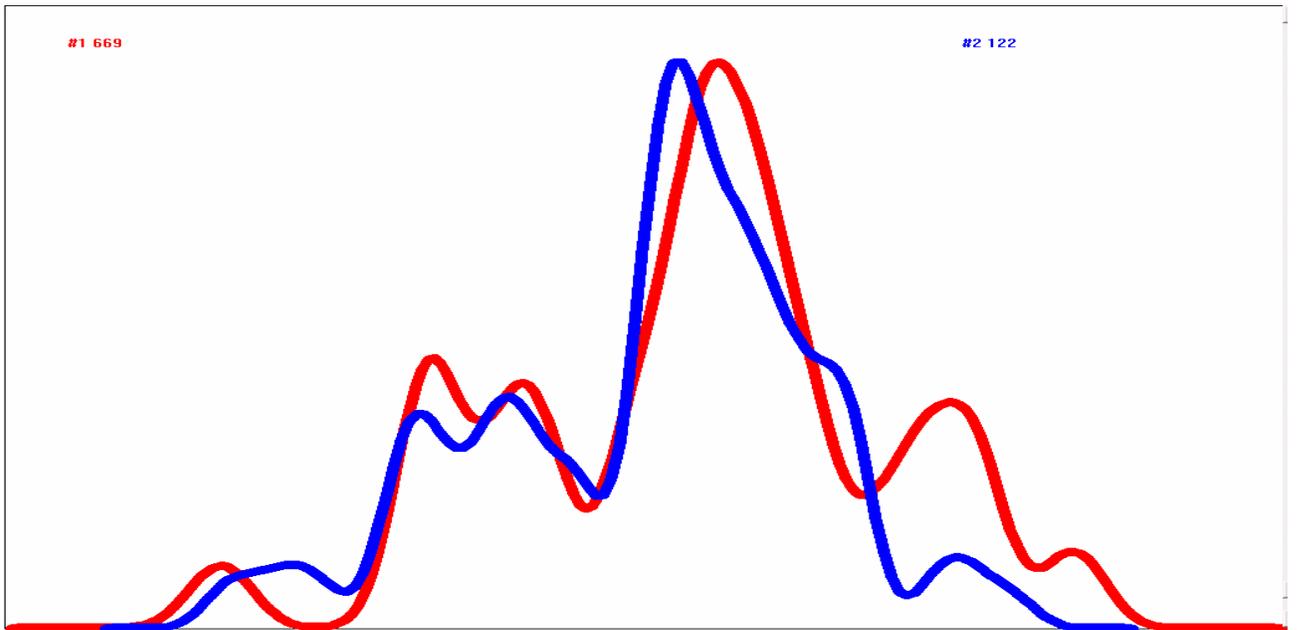

**Fig. 10** A similarity of the New-Moon histograms **at different geographic points at the same time** (Pushchino, the New Moon on 21 June 2001, histogram no. 669 and Antarctic, the New Moon on 21 June 2001, histogram no. 122). "Mirror" - the histograms which superimposed after turnover.

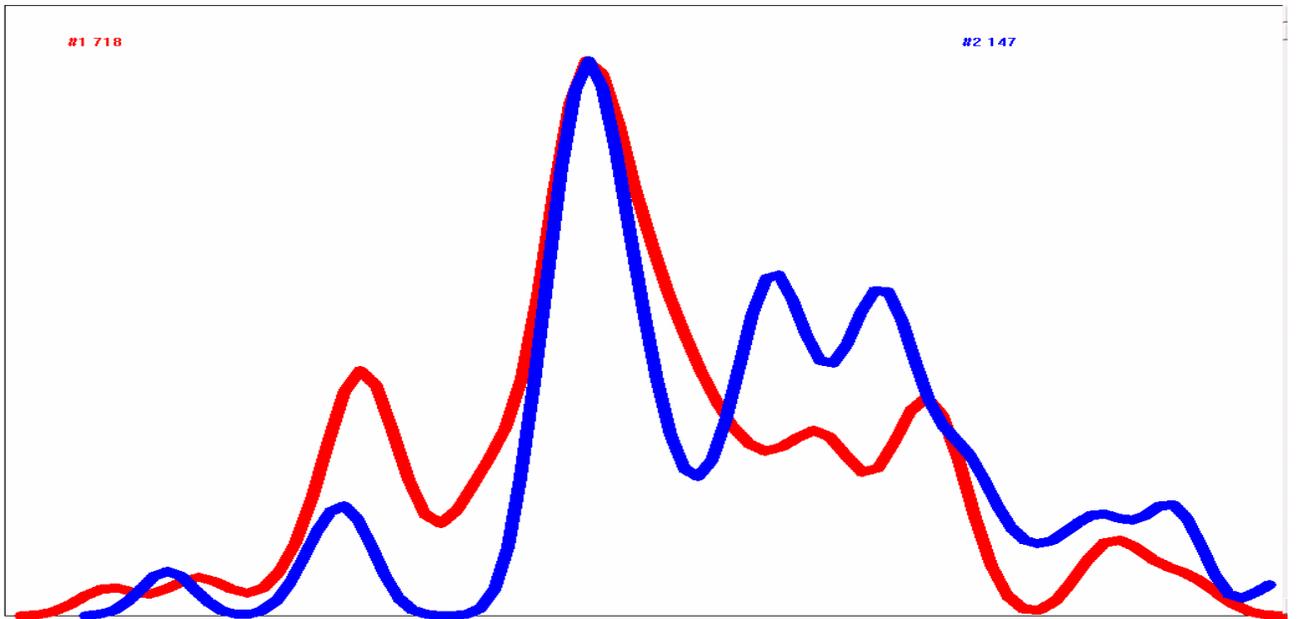

**Fig. 11** A similarity of the New-Moon histograms **at different geographic points at the same time** (Pushchino, the New Moon on 27 September 2000, histogram no. 147 and Arctic, the New Moon on 27 September 2000, histogram no. 718).

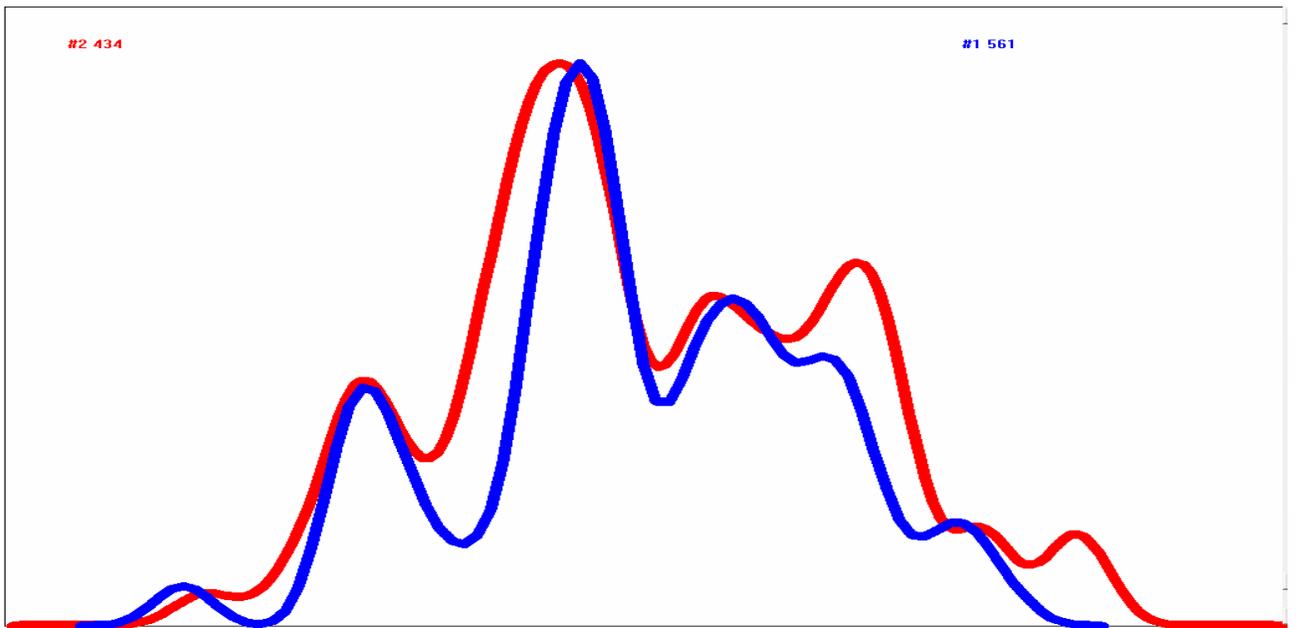

**Fig. 12** A similarity of the New-Moon histograms **at different geographic points at different times** (Pushchino, the New Moon on 17 September 2001, histogram no. 434 and Antarctic, the New Moon on 23 May 2001, no. 561).

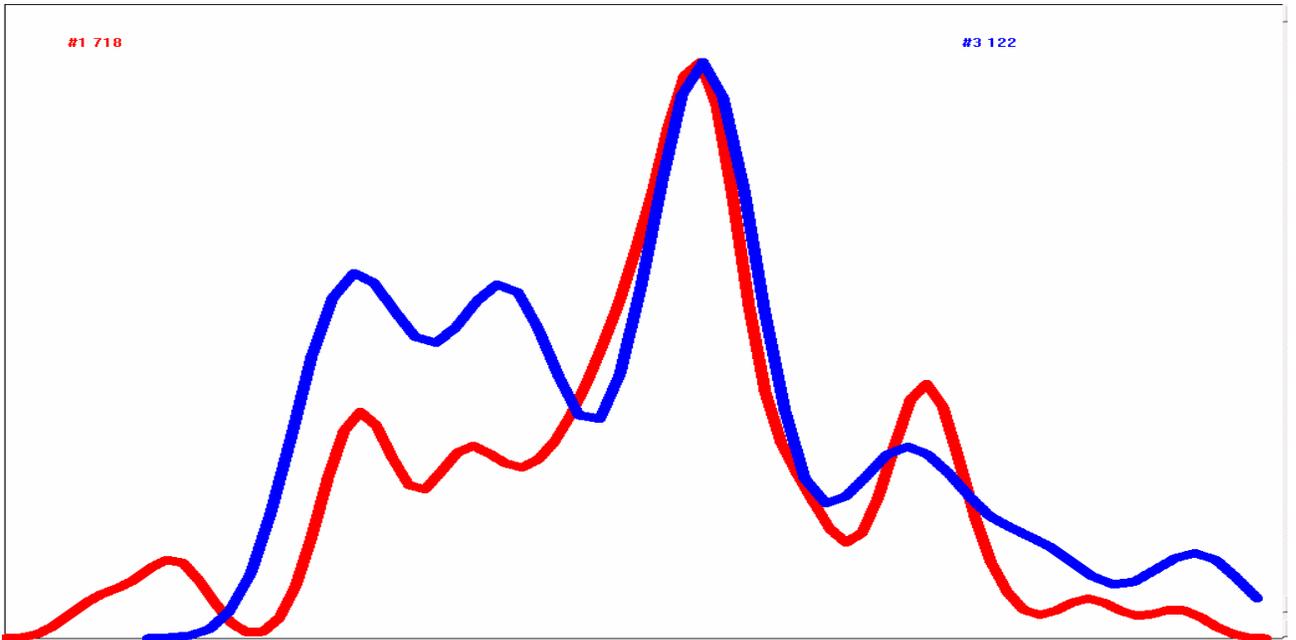

**Fig. 13** A similarity of the New-Moon histograms **at different geographic points at different times** (Arctic, the New Moon on 27 September 2000, histogram no. 718 and Antarctic, the New Moon on 21 June 2001, no. 122).

This effect is observed upon measurements of processes of different nature, including the measurements of α-radioactivity of $^{137}$**Cs** (in Dubna) and measurements of earth-crust oscillations near the Lake Baikal using a laser interferometer.

**Thus, at the New-Moon, simultaneously (to an accuracy of 2-10 min) in different geographic points throughout the Earth, the same characteristic fine structure of distribution is observed upon measurements of processes of different nature. It is of great importance that attempts to detect a similar effect during the Full Moon periods have not been successful.**

**3 Discussion**

*3.0.1 Is this phenomenon real?*

In all, we examined the form of histograms during 76 New-Moon periods. For 63 of them, histograms of a characteristic shape were observed. The probability of accidental occurrence of the histogram of a particular shape just in the new-moon period was estimated as follows. We determine the number of histograms (m) similar to a characteristic histogram in the entire massive of N histograms. Then the value of k (the difference between the calculated number of the histogram and the number of the histogram characteristic for the New Moon obtained in this case) is determined; e.g. the calculated number of the histogram must be 150, however, the histogram characteristic of the New Moon in this case has number 153 as it is shown in Fig. 2. Then the probability P of accidental occurrence of the histogram of a characteristic form ("New-Moon-form") in this geographical point will be:

$$P = m \cdot k/N. \qquad (1)$$

Given the time of constructing one histogram to be 2 min, the number of histograms that can be constructed during 24 h is 720. The number of histograms of the "New-Moon-form"(shown in red in a series of Fig. 1 - Fig. 8) varies in different experiments from 2 to 12 during 24 h. Hence, the probability of accidental occurrence of this particular form just in the New-Moon period in each particular case will be from $P_1 = 1.4 \cdot 10^{-2}$ to $P_2 = 8 \cdot 10^{-2}$ ($m_1 = 2$; $m_2 = 12$; $k = 5$; $N = 720$). The total probability of obtaining the same result upon repeating the measurements is equal to the product of probabilities calculated for individual experiments. Obviously, the probability of an accidental occurrence of the effect in several independent experiments is negligibly small. Thus, even for three experiments P is as low as $10^{-3}$. The probability of a random occurrence of this result can also be estimated using Fig. 1. For illustration, 21 histograms for each experiment are presented in the figure. As a rule, there is only one (more rarely two) characteristic histogram among them. Then the probability of obtaining a random result in each individual experiment is as low as $2.5 \cdot 10^{-1}$. Clearly, the probability of a random 8-fold recurrence of this result is extremely low (~$2 \cdot 10^{-7}$). Since this effect was observed in 63 out of 76 experiments during the New-Moon periods, the effect must be real.

*3.1 Possible explanation of the "New-Moon effect"*

Studies of the patterns in the occurrence of histograms of particular shapes led us to conclude that there is a universal external "reason" that determines the fine structure of distributions of the results of measurements of diverse processes. The only common feature of all the processes studied is that they occur in one and the same space-time. In view of this it was proposed that the regular changes (we call them as "macroscopic fluctuations") in the fine structure of histograms are due to corresponding changes (oscillations) in space-time "on a global scale" [7,10,11]. If it were so, it would be natural to assume that the original cause of these phenomena in general is a gravitational inhomogeneities (fluctuations), which become evident upon the rotation of the Earth around its axis (from whence "solar" and "sidereal" day; the occurrence of similar independent processes at the same local time) and the movement of the Earth along its circumsolar orbit (whence a yearly period). The periods of 27 and 29.5 days suggest that mutual positions of the Earth, the Moon, and the Sun must also be taken into account [4-13].

We believe that the fine structure of histograms results from the interference of some wave processes due to the influence of celestial bodies . The failure to observe similar effects in Full Moon periods suggests that "New-Moon effect" is due to the screening effect of the Sun on the Earth rather than to an increase in tidal forces. Virtually synchronous occurrence of a characteristic shape of histograms during the New-Moon periods at different geographical points separated by distances approximately equal to the Earth's diameter suggests that here we are dealing with the disturbances propagating with a relativistic velocity.

The authors are grateful to M.N. Kondrashova, L.A. Blumenfeld, V.N. Morozov for support and valuable discussion. We are indebted to V.A. Kolombet, N.V. Udaltzova, T.A. Zenchenko, A.A. Konradov, and E.V. Pozharskii for long-standing collaboration and valuable discussions. We thank E.V. Pozharskii for the development of a computer program for analyzing the forms of histograms. A special thank is to I.A. Rubinstein (Research Institute of Nuclear Physics, Moscow State University) for constructing a semiconductor detector for alpha-radioactivity measurements and to S.S. Zhirkov for the help in computer analysis of the results of measurements. We are indebted to V.P. Tikhonov and T. Peterson for valuable discussions and financial support.